\documentclass[10pt,a4paper]{article}

\usepackage{amsmath,amsgen,latexsym}
\usepackage{amstext,amssymb,amsfonts,latexsym}
\usepackage{theorem}
\usepackage{pifont}
\usepackage{microtype}

 \newcommand{\bs}{\bigskip}
 \newcommand{\ms}{\medskip}
 \newcommand{\n}{\noindent}
 \newcommand{\s}{\smallskip}
 \newcommand{\hs}[1]{\hspace*{ #1 mm}}
 \newcommand{\vs}[1]{\vspace*{ #1 mm}}

 \newcommand{\setempty}{\mathrm{\O}}
 
 \newcommand{\nat}{\mathbb{N}}

 \newcommand{\co}{\mathrm{co}\mbox{-}}

 \newcommand{\CC}{{\cal C}}

 \newcommand{\dl}{\mathrm{L}}
 \newcommand{\nl}{\mathrm{NL}}
 \newcommand{\p}{\mathrm{P}}
 \newcommand{\np}{\mathrm{NP}}

\theoremstyle{plain}
\theoremheaderfont{\bfseries}
 \newtheorem{theorem}{Theorem}[section]

 {\theorembodyfont{\rmfamily}
  \newtheorem{definition}[theorem]{Definition}}
 {\theorembodyfont{\rmfamily} }
 {\theorembodyfont{\rmfamily} }

 \newenvironment{proofof}[1]{\vspace*{5mm} \par \noindent
         {\bf Proof of #1.\hs{2}}}{\hfill$\Box$ \vspace*{3mm}}

 \newcommand{\ceilings}[1]{\lceil #1 \rceil}

\newcommand{\ignore}[1]{}

\newcommand{\cent}{{|}\!\!\mathrm{c}}
\newcommand{\dollar}{\$}

\newcommand{\psublin}{\mathrm{PsubLIN}}

\newcommand{\dstcon}{\mathrm{DSTCON}}

\newcommand{\para}{\mathrm{para}\mbox{-}}
\newcommand{\dtimespace}{\mathrm{DTIME},\mathrm{SPACE}}
\newcommand{\ntimespace}{\mathrm{NTIME},\mathrm{SPACE}}

\begin{document}

\pagestyle{plain}
\setcounter{page}{1}

\begin{center}
{\Large {\bf Supportive Oracles for Parameterized Polynomial-Time \s\\ Sub-Linear-Space Computations in Relation to L, NL, and P}}\bs\ms\\

{\sc Tomoyuki Yamakami}\footnote{Affiliation: Faculty of Engineering, University of Fukui, 3-9-1 Bunkyo, Fukui 910-8507,  Japan}\ms\\
\end{center}


\begin{quote}
\n{\bf Abstract:}
We focus our attention onto polynomial-time sub-linear-space computation for decision problems, which are parameterized by size parameters $m(x)$, where the informal term ``sub linear'' means a function of the form  $m(x)^{\varepsilon}\cdot polylog(|x|)$ on input instances $x$ for a certain absolute constant $\varepsilon\in(0,1)$ and a certain polylogarithmic function  $polylog(n)$. The parameterized complexity class $\psublin$ consists of all parameterized decision problems solvable simultaneously in polynomial time using sub-linear space. This complexity class is associated with the linear space hypothesis. There is no known inclusion relationships between $\psublin$ and $\para\nl$ (nondeterministic log-space class), where the prefix ``para-'' indicates the natural parameterization of a given complexity class.
Toward circumstantial evidences for the  inclusions and separations of the associated complexity classes, we seek their relativizations. However, the standard relativization of Turing machines is known to violate the relationships of $\dl\subseteq\nl=\co\nl\subseteq \mathrm{DSPACE}[O(\log^2{n})]\cap \p$. We instead consider special oracles, called $\nl$-supportive oracles, which guarantee these relationships in the corresponding relativized worlds. This paper vigorously constructs such NL-supportive oracles that generate relativized worlds where, for example, $\para\dl\neq \para\nl\nsubseteq \psublin$ and $\para\dl\neq \para\nl\subseteq \psublin$.

\s

\n{\bf Keywords:} {supportive oracle, parameterized decision problem, relativization, sub-linear space computation, log-space computation}
\end{quote}

\sloppy
\section{Prelude: Quick Overview}

\subsection{Size Parameters and Parameterized Decision Problems}\label{sec:size-parameter}

Among decision problems computable in polynomial time, nondeterministic logarithmic-space (or log-space, for short) computable problems are of special interest, partly because these problems contain practical problems, such as the \emph{directed $s$-$t$ connectivity problem} ($\dstcon$) and the \emph{2-CNF Boolean formula satisfiability problem} ($\mathrm{2SAT}$). Those problems form a complexity class known as $\nl$ (nondeterministic log-space class).

For many problems, their computational complexity have been discussed, from a more practical aspect, according to the ``size''  of particular items of each given instance. As a concrete example of such a ``size,'' let us consider  an efficient algorithm of Barnes, Buss, Ruzzo, and Schieber \cite{BBRS98} that solves DSTCON on input graphs of $n$ vertices and $m$ edges simultaneously using $(m+n)^{O(1)}$ time and $n^{1-c/\sqrt{\log{n}}}$ space for an appropriately chosen constant $c>0$. In this case, the number of vertices and the number of edges in a directed graph $G$ are treated as the ``size'' or the ``basis unit'' of measuring the computational complexity of DSTCON. For an input CNF Boolean formula, in contrast, we can take the number of variables and the number of clauses as the ``size'' of the Boolean formula. In a more general fashion, we denote the ``size'' of instance $x$ by $m(x)$ and we call this function $m$ a \emph{size parameter} of a decision problem $L$. A decision problem $L$ together with a size parameter $m$ naturally forms a \emph{parameterized decision problem} and we use a special notation $(L,m)$ to describe it.

Throughout this paper, we intend to study the properties of parameterized decision problems and their collections. Such collections are distinctively called \emph{parameterized complexity classes}. To distinguish such parameterized complexity classes from standard binary-size complexity classes, we often append the term ``para-'' as in $\para\nl$ and $\para\dl$, which are respectively the parameterizations of $\nl$ and $\dl$ (see Section \ref{sec:preliminary} for their formal definitions).

The aforementioned parameterized decision problems $(\dstcon,m_{ver})$ and $(2\mathrm{SAT},m_{vbl})$, where $m_{ver}(\cdot)$ and $m_{vbl}(\cdot)$ respectively indicate the number of vertices and the number of variables, fall into $\para\nl$ \cite{Yam17a}. It is, however, unclear whether we can improve the performance of the aforementioned algorithm of Barnes et al. to run using only $O(m_{ver}(x)^{\varepsilon}\ell(|x|))$ space for a certain absolute constant $\varepsilon\in[0,1)$ and a certain polylogarithmic (or polylog, for short) function $\ell$.  Given a size parameter $m(\cdot)$, the informal term ``sub linear'' generally refers to a function of the form $m(x)^{\varepsilon}\cdot\ell(|x|)$ for a certain constant $\varepsilon\in(0,1)$ and a certain polylog function $\ell$.
We denote by $\psublin$ the collection of all parameterized decision problems  solved by deterministic Turing machines running simultaneously in $(|x|m(x))^{O(1)}$ (polynomial) time using $O(m(x)^{\varepsilon}polylog(|x|))$ (sub-linear) space \cite{Yam17a}. It follows that $\para\dl\subseteq \psublin\subseteq \para\p$. Various sub-linear reducibilities was further studied in \cite{Yam17b} in association with $\psublin$.
The \emph{linear space hypothesis} (LSH), which is a practical working hypothesis proposed in \cite{Yam17a}, asserts that $(2\mathrm{SAT}_3,m_{vbl})$ cannot belong to $\psublin$, where $2\mathrm{SAT}_3$ is a variant of $2\mathrm{SAT}$ with an extra restriction that every 2-CNF Boolean formula given as an instance must have each variable appearing at most 3 times in the form of literals. We do not know whether LSH is true or even $\para\nl\nsubseteq\psublin$. A characterization of LSH was given in \cite{Yam18} in connection to state complexity of finite automata.

\subsection{Relativizations of $\dl$, $\nl$, $\p$, and $\psublin$}

The current knowledge seems not good enough to determine the exact complexity of $\psublin$ in comparison with $\para\dl$, $\para\nl$, and $\para\p$.
This fact makes us look for \emph{relativizations} of these classes by way of forcing underlying Turing machines to make queries to appropriately chosen oracles. The notion of relativization in computational complexity theory dates back to an early work of Baker, Gill, and Solovay \cite{BGS75}, who constructed various  relativized worlds in which various inclusion and separation  relationships among $\p$, $\np$, and $\co\np$ are possible.
Generally speaking, relativization is a methodology by which we can argue that a certain mathematical property holds or does not hold in the presence of external information source, called an \emph{oracle}. The use of an oracle $A$ creates a desired relativized world where certain desired conditions, such as   $\p^A\neq\np^A$ together with $\np^A=\co\np^A$, hold.
In a similar vein, we want to discuss the possibility/impossibility of the inclusion of $\para\nl$ in $\psublin$ by considering relativized worlds where various conflicting relationships between $\para\nl$ and $\psublin$ hold.

Unlike $\p$ and $\np$, it has been known that there is a glitch in defining the relativization of $\nl$. Ladner and Lynch \cite{LL76} first considered relativization of $\nl$ in a way similar to that of Baker, Gill, and Solovay \cite{BGS75}. Despite our knowledge regarding basic relationships among $\nl$, $\co\nl$, and $\p$, this relativization leads to the existence of oracles $A$ and $B$ such that $\nl^A\nsubseteq\p^A$ and $\nl^B\neq\co\nl^B$ although $\co\nl=\nl\subseteq \p$ holds in the unrelativized world.
Quite different from time-bounded oracle Turing machines, there have been several suggested models for space-bounded oracle Turing machines.
Ladner-Lynch relativization does not always guarantee both relationships $\nl^A\subseteq \p^A$ and $\nl^B=\co\nl^B$ since certain oracles $A,B$ refute those relations. This looks like contradicting the fact that, in the un-relativised world,  $\dl\subseteq \nl\subseteq \p$, $\nl=\co\nl$, and $\nl\subseteq\mathrm{LOG}^2\mathrm{SPACE}$ hold \cite{Imm88,Rei08,Sav70,Sze88}, where $\mathrm{LOG}^2\mathrm{SPACE}= \mathrm{DSPACE}[O(\log^2{n})]$.

Another, more restrictive relativization model was proposed in 1984 by Ruzzo, Simon, and Tompa \cite{RST84}, where an oracle machine behaves deterministically while writing a query word on its query tape. More precisely, after a query tape becomes blank, if the oracle machine starts writing the first symbol of a query word, then the machine must make deterministic moves until the query word is completed and an oracle is called. After the oracle answers, the query tape is automatically erased to be blank again and a tape head instantly jumps back to the initial tape cell. This restrictive model can guarantee the inclusions $\dl^A\subseteq \nl^A \subseteq\p^A$ for any oracle $A$; however,
it is too restrictive because it leads to the conclusion that $\dl=\nl$ iff $\dl^A=\nl^A$ for any oracle $A$ \cite{KL87}.

In this paper, we expect our relativization of an underlying machine to guarantee that all log-space nondeterministic oracle Turing machines can be simulated by polynomial-time deterministic Turing machines, yielding three relationships that $\para\nl^A\subseteq \para\p^A$, $\para\nl^A= \co\para\nl^A$, and $\para\nl^A\subseteq \para\mathrm{LOG}^2\mathrm{SPACE}^A$, where $\para\mathrm{LOG}^2\mathrm{SPACE}$ is the parameterization of $\mathrm{LOG}^2\mathrm{SPACE}$ and $\co\para\nl$ is the collection of all  $(L,m)$ for which $(\overline{L},m)$ belongs to $\para\nl$.

In spite of an amount of criticism, relativization remains an important research subject to pursue. Returning to parameterized complexity classes, nonetheless, it is possible to consider ``conditional'' relativizations that support the aforementioned three relations.
To distinguish such relativization from the ones we have discussed so far, we need a new type of relativization, which we will explain in the next subsection. In Section \ref{sec:discussion}, we will return to a discussion on the usefulness (and the vindication) of this new relativization.

\subsection{Main Contributions}\label{sec:introduction}

We use the notation $\psublin^A$ to denote the collection of all parameterized decision problems solvable by oracle Turing machines with adaptive access to oracle $A$ simultaneously using $(|x|m(x))^{O(1)}$ time and  $O(m(x)^{\varepsilon}\ell(|x|))$ space on all inputs $x$ for a certain constant $\varepsilon\in[0,1)$ and a certain polylog function $\ell$.

A key concept of our subsequent discussion is ``supportive oracles.''
Instead of restricting the way of accessing oracles (such as non-adaptive queries and limited number of queries), we use the most natural query mechanism but we force the oracles to support certain known inclusion relationships among complexity classes.
In this way, an oracle $A$ is said to be \emph{$\nl$-supportive} if the following three relations hold:  (1) $\para\dl^A\subseteq \para\nl^A\subseteq \para\p^A$, (2) $\para\nl^A= \co\para\nl^A$, and (3) $\para\nl^A \subseteq \para\mathrm{LOG}^2\mathrm{SPACE}^A$. Note that Condition (1) is always satisfied for any oracle $A$.

We first claim the existence of recursive $\nl$-supportive oracles generating various relativized worlds where three classes $\para\dl$, $\psublin$, and $\para\p$ have specific computational power. Notice that $\para\dl^A\subseteq \psublin^A \subseteq \para\p^A$ holds for all oracles $A$.

\begin{theorem}\label{oracle-L-and-PsubLIN}
There exist recursive $\nl$-supportive oracles $A$, $B$, $C$, and $D$ satisfying the following conditions.
\begin{enumerate}\vs{-1}
  \setlength{\topsep}{-2mm}%
  \setlength{\itemsep}{1mm}%
  \setlength{\parskip}{0cm}%

\item $\para\dl^A = \psublin^A = \para\p^A$.

\item $\para\dl^{B}\subsetneqq \psublin^{B}\subsetneqq \para\p^{B}$.

\item $\para\dl^{D} = \psublin^{D}\subsetneqq \para\p^{D}$.

\item $\para\dl^{C}\subsetneqq \psublin^{C}= \para\p^{C}$.
\end{enumerate}
\end{theorem}

The difficulty in proving each claim in Theorem \ref{oracle-L-and-PsubLIN} lies in the fact that we need to (i) deal with the fluctuations of the values of size parameters of parameterized decision problems (notice that the standard binary length of inputs is monotonically increasing) and to (ii) satisfy three or four conditions simultaneously for parameterized complexity classes by avoiding any conflict occurring during the construction of the desired oracles.

We can prove other relationships among $\para\dl$, $\para\nl$, and $\psublin$. Concerning a question of whether or not $\para\nl\subseteq \psublin$, we can present four different relativized worlds in which $\para\nl\subseteq \psublin$ and $\para\nl\nsubseteq \psublin$ separately hold between $\para\nl$ and $\psublin$ in relation to $\para\dl$.

\begin{theorem}\label{oracle-NL-PsubLIN}
There exist recursive $\nl$-supportive oracles $A$, $B$, $C$,  and $D$ satisfying the following conditions.
\begin{enumerate}\vs{-1}
  \setlength{\topsep}{-2mm}%
  \setlength{\itemsep}{1mm}%
  \setlength{\parskip}{0cm}%

\item $\para\dl^A = \para\nl^A = \psublin^A$.

\item $\para\dl^B = \para\nl^B\subsetneqq \psublin^B$.

\item $\para\dl^C\neq \para\nl^C\nsubseteq \psublin^C$.

\item $\para\dl^D\neq \para\nl^D\subseteq \psublin^D$.
\end{enumerate}
\end{theorem}

The relationships given in Theorems \ref{oracle-L-and-PsubLIN}--\ref{oracle-NL-PsubLIN} suggest that any relativizable proof is not sufficient to separate $\para\dl$, $\para\nl$, $\psublin$, and $\para\p$.

\section{Preliminaries}\label{sec:preliminary}

We briefly explain basic terminology necessary for the later sections. We use $\nat$ to denote the set of all \emph{natural numbers} (i.e., nonnegative integers) and we set $\nat^{+}=\nat-\{0\}$. Given a number $n\in\nat^{+}$, $[n]$ expresses the set $\{1,2,\ldots,n\}$. In this paper, all \emph{polynomials} have nonnegative integer coefficients and all \emph{logarithms} are taken to the base $2$. We define $\log^*{n}$ as follows. First, we set $\log^{(0)}{n}=\log{n}$ and $\log^{(i+1)}{n} = \log(\log^{(i)}{n})$ for each index $i\in\nat$. Finally, we set $\log^*{n}$ to be the minimal number $k\in\nat$ satisfying $\log^{(k)}{n}\leq1$.

An \emph{alphabet} $\Sigma$ is a nonempty finite set and a \emph{string over $\Sigma$} is a finite sequence of elements of $\Sigma$. A \emph{language over}  $\Sigma$ is a subset of $\Sigma^*$. We freely identify a decision problem with its associated language over $\Sigma$. The \emph{length} (or \emph{size}) of a string $x$ is the total number of symbols in $x$ and is denoted $|x|$. We write $\overline{L}$ for the set $\Sigma^*-L$ when $\Sigma$ is clear from the context. A function $f:\Sigma^*\to\Sigma^*$ (resp., $f:\Sigma^*\to\nat$) is \emph{polynomially bounded} if there exists a polynomial $p$ satisfying $|f(x)|\leq p(|x|)$ (resp., $f(x)\leq p(|x|)$) for all $x\in\Sigma^*$.

We use \emph{deterministic Turing machines} (DTMs) and \emph{nondeterministic Turing machines} (NTMs), each of which  has a read-only input tape and a rewritable work tape. If necessary, we also attach a write-only\footnote{A tape is \emph{write only} if a tape head must move to the right whenever it write any non-blank symbol.} output tape. All tapes have the left endmarker $\cent$ and stretch to the right. Additionally, an input tape has the right endmarker $\dollar$.
When a DTM begins with an initial state and, whenever it enters a halting state (either an accepting state or a rejecting state), it halts.
We say that a DTM $M$ \emph{accepts} (resp., \emph{rejects}) input $x$ if $M$ starts with $x$ written on an input tape (surrounded by the two endmarkers) and  enters an accepting (resp., a rejecting) state. Similarly, an NTM \emph{accepts} $x$ if there exists a series of nondeterministic choices that lead the NTM to an accepting state. Otherwise, the NTM \emph{rejects} $x$.
A machine $M$ is said to \emph{recognize} a language $L$ if, for all $x\in L$, $M$ accepts $x$, and for all $x\in\Sigma^*- L$, $M$ rejects $x$.

The notation $\dl$ (resp., $\nl$) refers to the class of all languages recognized by DTMs (resp., NTMs) using space $O(\log{n})$, where ``$n$'' is a symbolic input size. Moreover, $\p$ stands for the class of languages recognized by DTMs in time $n^{O(1)}$. It is known that $\dl\subseteq \nl=\co\nl\subseteq \mathrm{LOG}^2\mathrm{SPACE}\cap\p$ \cite{Imm88,Rei08,Sav70,Sze88}.

An \emph{oracle} is an external device that provides useful information to an underlying Turing machine, which is known as an \emph{oracle Turing machine}. In this paper, oracles are simply languages over a certain alphabet.
An oracle Turing machine $M$ is equipped with an extra \emph{query tape}, in which the machine writes a query word, say, $w$ and enters a query state $q_{query}$ that triggers an oracle query. We demand that any query tape should be \emph{write only} because, otherwise, the query tape can be used as an extra work tape composed of polynomially many tape cells. Triggered by an oracle query, an oracle $X$ responds by modifying the machine's inner state from $q_{query}$ to either $q_{yes}$ or $q_{no}$, depending on $w\in X$ or $w\notin X$, respectively. Simultaneously, the query tape becomes empty and its tape head is returned to $\cent$. Given an oracle Turing machine $M$ and an oracle $A$, the notation $L(M,A)$ expresses the set of all strings accepted by $M$ relative to $A$.

A \emph{size parameter} is a function from $\Sigma^*$ to $\nat^{+}$ for a certain alphabet $\Sigma$. A \emph{log-space size parameter} $m:\Sigma^*\to\nat$ is a size parameter for which there exists a DTM $M$ equipped with a write-only output tape that takes a string $x\in\Sigma^*$ and produces $1^{m(x)}$ on the output tape using $O(\log|x|)$ space. As a special size parameter, we write ``$||$'' to denote the size parameter $m$ defined by $m(x)=|x|$ for any $x$. The notation $\mathrm{LSP}$ indicates the set of all log-space size parameters. Given a size parameter $m$ and any index $n\in\nat$, we set $\Sigma_n=\{x\in\Sigma^*\mid m(x)=n\}$. Note that $\Sigma_i\cap\Sigma_j=\setempty$ for any distinct pair $i,j\in\nat$ and that $\Sigma^*=\bigcup_{n\in\nat}\Sigma_n$.
A pair $(L,m)$ with a decision problem (equivalently, a language) $L$ and a size parameter $m$ is called a \emph{parameterized decision problem} and any collection of parameterized decision problems is called a \emph{parameterized
complexity class}. We informally use the term ``parameterization'' for underlying decision problems and complexity classes if we supplement size parameters  to their instances.

As noted in Section \ref{sec:size-parameter}, the prefix ``para-'' is used to  distinguish parameterized complexity classes
from standard complexity classes. With this convention, for two functions $s$ and $t$, the notation $\para\dtimespace(t(|x|,m(x)),s(|x|,m(x)))$, where ``$m(x)$'' in ``$\log{m(x)}$'' indicates a symbolic size parameter $m$ with a symbolic input $x$, denotes the collection of all parameterized decision problems with log-space size parameters, each $(L,m)$ of which is solved (or recognized) by a certain DTM $M$ in time $O(t(|x|,m(x)))$ using space $O(s(|x|,m(x)))$. Its nondeterministic variant is denoted by $\para\ntimespace(t(|x|,m(x)),s(|x|,m(x)))$.
We set $\para\nl$ to be $\bigcup_{c\in\nat} \para\ntimespace((|x|m(x))^c,\log{|x|m(x)})$ and $\para\dl$ to be $\bigcup_{c\in\nat} \dtimespace((|x|m(x))^c,\log{|x|m(x)})$. Moreover, we set    $\para\mathrm{LOG}^2\mathrm{SPACE}$ to be $\para\dtimespace(|x|^{\log|x|}m(x)^{\log{m(x)}},\log^2|x|+\log^2{m(x)})$.
When we take $m(x)=|x|$, those parameterized complexity classes coincide with the corresponding ``standard'' complexity classes.
In addition, we define $\psublin$ as $\bigcup_{c,k\in\nat,\varepsilon\in[0,1)} \dtimespace((|x|m(x))^c,m(x)^{\varepsilon}\log^k|x|)$.
Given a parameterized complexity class $\para\CC$, its complement class $\co\para\CC$ is composed of all parameterized decision problems $(L,m)$ for which $(L,m)$ belongs to $\para\CC$. The relativization of $\para\nl$ with an oracle $A$ is denoted by $\para\nl^A$ and is obtained by replacing underlying Turing machines for $\para\nl$ with oracle Turing machines. In a similar fashion, we define $\para\dl^A$, $\para\p^A$, and $\psublin^A$.

Formally, we introduce the notion of $\nl$-supportive oracles.

\begin{definition}
An oracle $A$ is said to be \emph{$\nl$-supportive} if the following three conditions hold:  (1) $\para\dl^A\subseteq \para\nl^A\subseteq \para\p^A$, (2) $\para\nl^A= \co\para\nl^A$, and (3) $\para\nl^A \subseteq \para\mathrm{LOG}^2\mathrm{SPACE}^A$.
\end{definition}

Although Condition (1) holds for all oracles $A$, we include it for a clarity reason.

In the subsequence sections, we will provide the proofs of Theorems \ref{oracle-L-and-PsubLIN}--\ref{oracle-NL-PsubLIN}.

\section{Proofs of Theorem \ref{oracle-L-and-PsubLIN}}\label{sec:proof-of-Theorem-1}

We will give necessary proofs that verify our main theorems.
We begin with proving Theorem \ref{oracle-L-and-PsubLIN}.

Our goal is to construct oracles $A$, $B$, $C$, and $D$ that satisfy Theorem \ref{oracle-L-and-PsubLIN}(1)--(4). Here, we start with the first claim of Theorem \ref{oracle-L-and-PsubLIN}.

\begin{proofof}{(1)}
Note that, if $\para\dl^A=\para\p^A$, then $A$ is $\nl$-supportive because we obtain $\para\nl^A =\co\para\nl^A = \para\dl^A  \subseteq\para\mathrm{LOG}^2\mathrm{SPACE}$. Let $A$ be any $\p$-complete problem (via log-space many-one reductions). We first claim that $\para\p^A\subseteq \para\dl^A$. Since $A\in\p$, we obtain $\para\p^A\subseteq \para\p$. Since $A$ is $\p$-complete, it follows that $\para\p\subseteq \para\dl^A$, as requested.
\end{proofof}

In the proofs of Theorem \ref{oracle-L-and-PsubLIN}(2)--(4) that will follow shortly, we need several effective enumerations of pairs consisting of machines and size parameters. First, let $\{(M_i,m_i)\}_{i\in\nat^{+}}$ be  an effective enumeration of all such pairs satisfying that, for each index $i\in\nat^{+}$,  $m_i$ is in $LSP$ and $M_i$ is a DTM running in time at most $(|x|m_i(x))^{c_i}+c_i$ using space at most $m_i(x)^{\varepsilon_i}\log^{k_i}{|x|}+c_i$ on all inputs $x$ and for all oracles for appropriately chosen constants $k_i,e_i>0$ and $\varepsilon_i\in[0,1)$. Moreover, let $\{(D_i,m_i)\}_{i\in\nat^{+}}$ denote an effective enumeration of pairs for which each $m_i$ belongs to $LSP$ and each $D_i$ is a DTM running in time at most $(|x|m_i(x))^{a_i}+a_i$ using space at most $a_i\log{|x|m_i(x)}+a_i$ on all inputs $x$ and for all oracles for a certain constant $a_i>0$.
Next, we use an effective enumeration $\{(P_i,m_i)\}_{i\in\nat^{+}}$, where each  $m_i$ is in $LSP$ and each DTMs $P_i$ runs in time at most $(|x|m_i(x))^{b_i}+b_i$ on all inputs $x$ and for all oracles, where $b_i$ is an absolute positive constant.
We also assume an effective enumeration $\{(N_i,m_i)\}_{i\in\nat^{+}}$ such that, for each $i\in\nat^{+}$, $m_i$ is in $LSP$ and $N_i$ is an NTM running in time at most $(|x|m_i(x))^{e_i}+e_i$ using space at most $e_i\log{|x|m_i(x)}+e_i$ on all inputs $x$ and for all oracles for a certain absolute constant $e_i>0$.  For notational simplicity, we write $\overline{M}$ to express an oracle machine obtained from $M$ by exchanging between $Q_{acc}$ and $Q_{rej}$. With this notation, each $\overline{N}_i$ relative to oracle $A$ together with $m_i$ induces a parameterized decision problem in $\co\para\nl^A$.

Since every log-space size parameter is computed by a certain log-space DTM equipped with an output tape, we can enumerate all log-space size parameters by listing all such DTMs as $(K_1,K_2,\ldots)$. For each index $i\in\nat^{+}$, we write $m_j$ for the size parameter computed by $K_j$ as long as it is obvious from the context. Since $m_j$ is polynomially bounded, it is possible to assume that $m_j(x)\leq |x|^{g_j}+g_j$ for all $j$ and $x$, where $g_j$ is an absolute positive constant.

Our construction of the desired oracles will proceed by stages. To prepare such stages, for a given finite set $\Theta\subseteq \nat^{+}$, let us define an index set $\Lambda=\{(n,l)\mid l\in\Theta,n\in\nat^{+} \}\cup\{0\}$ together with an  appropriate effective enumeration of all elements in $\Lambda$ defined by the following linear order $<$ on $\Lambda$:  (1)  $0<t$ holds for all $t\in\Lambda-\{0\}$ and (2) $(n',l')>(n,l)$ iff either $n'>n$ or
$n'=n\wedge l'>l$. Given a number $n\in\nat^{+}$, let $S_n = \{(x,i)\mid x\in\Sigma^n,i\in[\log^*{n}]\}$. Note that $|S_n|=2^n\log^*{n}$. We also define a linear ordering $<$ on $S_n$ with respect to $\{e_i\}_{i\in\nat^{+}}$ in the following way: letting $k_{x,i} = (|x|m_i(x))^{e_i}+e_i$, $(x,i)<(y,j)$ iff one of the following conditions hold: $k_{x,i} < k_{y,j}$, $k_{x,i}= k_{y,j}\wedge i< j$, and $k_{x,i}=k_{y,j}\wedge i=j \wedge x<y$ (lexicographically). According to this ordering $<$, we choose all elements of $S_n$ one by one
in the increasing order.

\begin{proofof}{(2)}
We wish to construct an oracle $D$ that meets the following four conditions: (i) $\psublin^B\nsubseteq \para\dl^B$, (ii) $\para\p^B\nsubseteq \psublin^B$, (iii) $\para\nl^B\subseteq \para\mathrm{LOG}^2\mathrm{SPACE}^B$, and (iv) $\co\para\nl^B \subseteq \para\nl^B$. These conditions obviously ensure the desired claim of $\para\dl^B\subsetneqq \psublin^B\subsetneqq \para\p^B$.

We want to introduce two example languages for (i) and (ii). First, we set $y_j = 10^{r_x-j}1^{j}$ and $u_j=B(101^ix\#0y_j)$ for all $j\in[r_x]$, where  $r_x=\ceilings{\sqrt{|x|}}$. We write  $u$ for $u_1u_2\cdots u_{r_x}$ and set $L_1^B=\{101^ix \mid 101^ix\#1u\in B\}$. It is not difficult to show that $(L_1^B,||)$ belongs to $\psublin^B$ for any oracle $B$ by running the following algorithm: first produce all words $101^ix\# 0y_j$ one by one, query them to obtain $u$ from $B$, remember all answer bits $u_j$, and finally query $101^ix\# 1u$. Similarly, we define  $y'_j=10^{|x|-j}1^j$ and $u_j=B(1^201^ix\#0y'_j)$ for each  $j\in[|x|]$ and set $L_2^B=\{1^201^ix \mid 1^201^ix\# 1u'\in B\}$, where $u'=u_1u_2\cdots u_n$. Note that $(L_2^B,||)\in\para\p^B$ for any oracle $B$. Through our oracle construction, we will define two sequences $\{n_{s_1}\}_{s_1\in\nat^{+}}$ and $\{n'_{s_2}\}_{s_2\in\nat^{+}}$. For readability, we set  $k_{x,i} =(|x|m_i(x))^{a_i}+a_i$, $k'_{x,i} = (|x|m_i(x))^{c_i}+c_i$, and $k''_{x,i} = (|x|m_i(x))^{e_i}+e_i$ for any $x\in\Sigma^*$ and $i\in\nat^{+}$.

In this proof, we set $\Theta=\{1,2,3\}$ and define $\Lambda$ as stated before. For each $t\in \Lambda$, we want to construct two sets $B_t$ and $R_t$. At Stage $0$, we set $B_0=R_0=\setempty$ and $n_0=n'_0=1$. Moreover, we set two counters $s_1$ and $s_2$ to $1$. In what follows, we deal with Stage $t=(n,l)\in\Lambda$ and the values of $s_1$ and $s_2$. By induction hypothesis, we assume that, for all $t'<t$, $B_{t'}$ and $R_{t'}$ have been already defined. Moreover, we assume that, for all $e_1<s_1$ and $e_2<s_2$, $n_e$ and $n'_{e'}$ have been appropriately defined. For simplicity, let $B'=\bigcup_{t'<t} B_{t'}$ and $R'=\bigcup_{t'<t}R_{t'}$.

During the construction process of $B$, the value of $k_{x,i}$ may fluctuate, depending on $(x,i)$,  and this fact may have many words reserved, leaving no room for inserting extra strings to define $B$ in (c)--(d). To avoid such a situation, we will use $S_n$ and its linear ordering $<$ with respect to $\{c_i\}_{i\in\nat^{+}}$. For our convenience, let $Z^{(3)}_{x,n} =\{1^301^ix\# 0^{k''_{x,i}}z\mid |z|=\ceilings{2\log\log{n}}\}$ and $Z^{(4)}_{x,n} =\{1^401^ix\# z\mid |z|=k''_{x,i}\}$.

\s

(a) Case $l=1$. Our target is Condition (i). Consider the size parameter $m(x)=|x|$.  If $n< \max\{n_{s_1},n'_{s_2}\}$, then we skip this case and move to Case $l=2$. Now, let us assume otherwise. We try to find a room for diagonalization by avoiding the reserved words set in the previous stages. For this purpose, we first  check whether there exist a number $\tilde{n}\in\nat^{+}$ and a string $x\in\Sigma^{\tilde{n}}$ satisfying that (*) $\max_{i\in[\log^*{n}]}\{4 n^{2a_ig_i}+a_i\} < \tilde{n} \leq 2^{n}$  and, for any $j\in\{3,4\}$, $|R'\cap Z^{(j)}_{x,\tilde{n}}| + |\tilde{x}|^{2a_i}+a_i + r_x +1 < \tilde{n}^{\log{\tilde{n}}}$, where $\tilde{x}=101^ix$.
The latter condition is to make enough room for $L_1^B$ as well as the constructions in (c)--(d). If (*) is not satisfied for all $\tilde{n}$ and $x$, then we skip this case. Assuming that (*) is satisfied for certain $\tilde{n}$ and $x$, we fix such a pair $(\tilde{n},x)$ for the subsequent argument.

Take the machine $D_i$ and the input $\tilde{x}=101^ix$. Recall that $D_i$ runs in time at most $|\tilde{x}|^{2a_i}+a_i$ using space at most $2a_i\log|\tilde{x}|+a_i$. Let $\tilde{R}_t  =\{101^ix\# 0y_j\mid j\in[r_x]\}$. Although $D_i$ may possibly query all words of the form $101^ix\#0y_j$ for $j\in[r_x]$, it cannot remember all oracle answers $u_j$, because the work tape space of $D_i$ is smaller than $r_x$. From this fact and also by (*), there is a set $B_t\subseteq \tilde{R}_t \cup\{101^ix\#1u\mid u=u_1\cdots u_{r_x}, u_j=B'(101^ix\# 0y_j), j\in[r_x]\} - R'$ for which $D_i^{B'\cup B_t}(\tilde{x}) \neq L_1^{B'\cup B_t}(\tilde{x})$. We define  $R_t$ to include $\tilde{R}_t\cup B_t$, all queried words of $D_i$ on $\tilde{x}$ relative to $B'\cup B_t$, and $101^ix\#1u$, where $u=u_1\cdots u_{r_x}$ and $u_j=B'(101^ix\# 0y_j)$ for all $j\in[r_x]$. Note that $|R'\cap R_t|< \tilde{n}^{\log{\tilde{n}}}$. Before leaving this case, we set $n_{s_1+1}$ to be $\tilde{n}$ and then increment the counter from $s_1$ to $s_1+1$.

\s

(b) Case $l=2$. Hereafter, we try to satisfy Condition (ii). For this purpose, let us assume that $B'$ and $R'$ have been updated. We will make an argument similar to (a) using $L_2^B$ instead of $L_1^B$. We skip this case and move to Case $l=3$ if $n< \max\{n_{s_1},n'_{s_2}\}$.  Otherwise, we check if there are a number $\tilde{n}$ and a string $x\in\Sigma^{\tilde{n}}$ satisfying that (*) $\max_{i\in[\log^*{n}]} \{4n^{2c_ig_i}+c_i\} < \tilde{n}\leq 2^{n}$ and $|R'\cap Z^{(j)}_{x,\tilde{n}}| + |\tilde{x}|^{2c_i}+c_i \tilde{n}+1 < \tilde{n}^{\log{\tilde{n}}}$ for any index $j\in\{3,4\}$, where $\tilde{x}=1^201^ix$. If no pair $(\tilde{n},x)$ satisfies (*), then we skip this case as well. Next, we assume (*) for certain $\tilde{n}$ and $x$.
We consider the machine $M_i$ and feed the input $\tilde{x}=1^201^ix$ to $M_i$.
We then choose a set $B_t\subseteq \tilde{R}_t \cup\{101^ix\#1u\mid u=u_1\cdots u_{\tilde{n}}, u_j=B'(1^201^ix\# 0y_j), j\in[\tilde{n}]\} - R'$, where $y_j = 1^{\tilde{n}-j}0^j$, satisfying $M_i^{B'\cup B_t}(\tilde{x}) \neq L_2^{B'\cup B_t}(\tilde{x})$. Note that $M_i$ cannot remember all values $u_j$ for any  $j\in[\tilde{n}]$ using its work tape because the work tape space is bounded by $|\tilde{x}|^{\varepsilon_i}\log^{k_{i}}|\tilde{x}|+c_i< \tilde{n}$.
Let $R_t$ be composed of $\tilde{R}_t$, all queried words of $M_i$ on $\tilde{x}$ relative to $B'\cup B_t$, and $1^201^ix\# 1u$, where $u=u_1\cdots u_{\tilde{n}}$ and $u_j = B'(1^201^ix\# 0y_j)$ for all $j\in[\tilde{n}]$. Finally, we set $n'_{s_2+1} = \tilde{n}$ and increment the counter from $s_2$ to $s_2+1$.

\s

(c) Case $l=3$. We target Condition (iii). Consider $S_n$ and its linear ordering $<$ with respect to $\{e_i\}_{i\in\nat^{+}}$. We inductively choose all pairs  $(x,i)$ in $S_n$ one by one in the increasing order. For each element $(x,i)$, let us consider $N_i$ with $x$ and $B'$. For convenience, we write $B_{t,<(x,i)}$ to denote the union of all $B_{t,(y,j)}$ for any $(y,j)\in S_n$ with $(y,j) <(x,i)$. Similarly, we write $R_{t,<(x,i)}$. In this case, we need to find an appropriate query word deterministically to simulate $N_i$ on $x$.
Since $m_i(x)\leq |x|^{g_i}+g_i$, it follows that $k''_{x,i} = (|x|m_i(x))^{e_i}+e_i \leq 4|x|^{e_ig_i}+e_i$. We set $W_{x,i} = \{1^301^ix\#0^{k''_{x,i}} y\mid |y|=\ceilings{2\log\log{|x|}}\}$.

Here, we define a new machine $H_i$ as follows. On input $w$, query all words of the form $1^301^iw\#0^{k''_{w,i}} y$ for any $y$ of length  $\ceilings{2\log\log|w|}$. Note that the number of different $y$'s is $2^{\ceilings{2\log\log|w|}}$, which is at most $2\log^2|w|$. Collect all answers from an oracle. Let $u$ be the sequence of oracle answers in order. Finally, make a query of $1^301^iw\#1^{k''_{w,i}} u$. If the oracle answers YES, accept $w$; otherwise, reject $w$.
Take strings of the form $1^301^ix\#0^{k''_{x,i}} y$ in $W_{x,i}$ so that, for the string $u$ obtained from $B'\cup B_{t,<(x,i)}$, $1^301^ix\#1^{k''_{x,i}} u$ is not in $R'\cup R_{t,<(x,i)}$. We include all those strings into $B_{t,(x,i)}$. It follows that $x\in L(N_i,B'\cup B_{t,<(x,i)}\cup B_{t,(x,i)})$ iff $x\in L(H_i, B'\cup B_{t,<(x,i)}\cup B_{t,(x,i)})$. Next, we define $R_{t,(x,i)}$ to include all queried strings of $N_i$, $\{1^301^ix\# 0^{k''_{x,i}} y\mid |y|=\ceilings{2\log\log{|x|}}\}$, and $1^301^ix\# 1^{k''_{x,i}} u$, where $u$ is the word determined by the query answers. In the end, we set $B_t = \bigcup_{(x,i)\in S_n} B_{t,(x,i)}$ and $R_t = \bigcup_{(x,i)\in S_n} R_{t,(x,i)}$.

\s

(d) Case $l=4$. We aim at Condition (iv). Consider $S_n$ and its linear ordering $<$ with respect to $\{c_i\}_{i\in\nat^{+}}$. Choose all pairs $(x,i)\in S_n$ one by one and define two sets $B_{t,(x,i)}$ and $R_{t,(x,i)}$. We run $\overline{N}_i$ on the input $x$ with the oracle $C'$. Note that $\overline{N}_i$ runs in time at most $k''_{x,i}$ using space at most $c_i\log{|x|m_i(x)}+c_i$ for all oracles. Note that $k''_{x,i}\leq 4|x|^{2c_ig_i}+c_i$ since $m_i(x)\leq |x|^{g_i}+g_i$. Let us consider the set $V_t= \{1^401^ix\# z\mid |z|=k''_{x,i}\}$. Note that the runtime bound of $N_i$ makes it impossible for $\overline{N}_i$ to  query any string in $V_t$.
A new machine $G_i$ is defined to work as follows: on input $w$, nondeterministically generate $1^401^iw\# z$ for all strings $z\in\Sigma^{k''_{w,i}}$ and query it. If an oracle answers YES, then accept $w$; otherwise, reject $w$.
If $(x,i)$ is the smallest element in $S_n$, then we set $B_{t,<(x,i)} = R_{t,<(x,i)}=\setempty$; otherwise, we define $R_{t,<(x,i)}=\bigcup_{(y,j)<(x,i)}R_{t,(y,j)}$.

We define $B_{t,(x,i)}$ as follows: if $\overline{N}_i$ accepts $x$ relative to $B'$, then we set $B_{t,(x,i)} =\{1^401^ix\#z_{x,i}\}$, where $z_{x,i} =\min\{z\in\Sigma^{k''_{x,i}}\mid 1^401^ix\#z\notin R'\cup R_{t,<(x,i)}\}$; otherwise, we set $C_{t,(x,i)}=\setempty$. We define $R_{t,(x,i)} = R' \cup \{w\mid \text{$w$ is queries by $\overline{N}_i$ on $x$}\}$. It follows by the definition that $x\in L(\overline{N}_i,B'\cup B_{t,<(x,i)} \cup B_{t,(x,i)})$ iff $x\notin L(G_i,B'\cup B_{t,<(x,i)} \cup B_{t,(x,i)})$. Before leaving this case,  we set $B_t = \bigcup_{(x,i)\in S_n} B_{t,(x,i)}$ and $R_t = \bigcup_{(x,i)\in S_n} R_{t,(x,i)}$.

\s

Finally, we define $B = \bigcup_{t\in\Lambda} B_t$. By the construction of $B$, Conditions (i)--(iv) are all satisfied.
\end{proofof}

The third claim of Theorem \ref{oracle-L-and-PsubLIN} is proven below.

\begin{proofof}{(3)}
To verify the target claim (3), it suffices for us to construct a set $C$ for which (i) $\para\p^C\nsubseteq \psublin^C$, (ii) $\para\nl^C\subseteq \para\dl^C$, and (iii) $\psublin^C\subseteq \para\dl^C$.

Similar to the proof of (2), we prepare $\Lambda$, $S_n$, $C_t$, and $R_t$ with a counter $s$, starting at $s=1$. Let us assume that we reach Stage $t=(n,l)\in\Lambda$ and the counter has advanced to $s$. Initially, we set $C'=\bigcup_{t'<t} C_{t'}$ and $R' = \bigcup_{t'<t} R_{t'}$. Case $l=1$, which targets Condition (i), is similar to that in the proof of (2). In what follows, we discuss only Cases $l\in\{2,3\}$. For simplicity, we set $k_{x,i} = (|x|m_i(x))^{e_i}+e_i$ and $k'_{x,i} = (|x|m_i(x))^{c_i}+c_i$ for any $x$ and $i$.

\s

(a) Case $l=2$. We aim at fulfilling Condition (ii). After treating Case $l=1$, we assume that $C'$ and $R'$ have been properly updated.
Using a linear ordering $<$ on $S_n$ with respect to $\{e_i\}_{i\in\nat^{+}}$, we choose all pairs $(x,i)$ in $S_n$ one by one in the increasing order. Consider the computation of $N_i$ on the input $x$ in time at most $k_{x,i}$ using space at most $e_i\log{|x|m_i(x)}+e_i$.
Let $R_{t,(x,i)}$ denote the set of all query words of $N_i$ on $x$ relative to $C'\cup C_{t,<(x,i)}$.
Here, we introduce a new DTM $E_i$ that works as follows. On input $w$, we make a query of the form $1^201^iw\# 0^{k_{w,i}}$ and accepts (resp., rejects) $w$ if its oracle answer is YES (resp., NO).
Define $C_{t,(x,i)} =\{1^201^ix\# 0^{k_{x,i}}\}$ if $N_i$ accepts $x$ relative to $B'\cup C_{t,<(x,i)}$, and $C_{t,(x,i)} =\setempty$ otherwise. Since $N_i$ cannot query $1^20^ix\# 0^{k_{x,i}}$, it follows that $x\in L(N_i,B'\cup C_{t,<(x,i)}\cup C_{t,(x,i)})$ iff $x\in L(E_i,C'\cup C_{t,<(x,i)}\cup C_{t,(x,i)})$.

\s

(b) Case $l=3$. Our goal is to meet Condition (iii). Here, we use a linear ordering $<$ on $S_n$ with respect to $\{c_i\}_{i\in\nat^{+}}$. Similarly to (a), we assume that $B'$ and $R'$ have been properly updated after Case $l=2$. Consider $M_i$ and pick up all pairs $(x,i)\in S_n$ one by one in the increasing order according to $<$.  Let us assume that we have already defined $B_{t,<(x,i)}$ and $R_{t,<(x,i)}$. Let $R_{t,(x,i)}$ be composed of all query words of $M_i$ on $x$ relative to $C'\cup C_{t,<(x,i)}$. Since $M_i$ runs in time at most $k'_{x,i}$, we define $C_{t,(x,i)} =\{1^301^ix\# 0^{k'_{x,i}}\}$ if $M_i$ accepts $x$ relative to $C'\cup C_{t,<(x,i)}$, and $C_{t,(x,i)}=\setempty$ otherwise.
Consider a new DTM $F_i$ defined as follows. On input $w$, compute $k'_{w,i}$, query $1^31^iw\# 0^{k'_{w,i}}$, accept (resp., reject) $w$ if the oracle answers  YES (resp., NO). We then obtain a relationship that $x\in L(M_i,C'\cup C_{t,<(x,i)}\cup C_{t,(x,i)})$ iff $x\in L(F_i,C'\cup C_{t,<(x,i)}\cup C_{t,(x,i)})$. Notice that $F_i$ uses only space $O(\log{k'_{x,i}})$.

\s

Finally, we set $C = \bigcup_{t\in\Lambda} C_t$. Clearly, Conditions (i)--(iii) are satisfied by this oracle $C$.
\end{proofof}

Next, we want to prove Theorem \ref{oracle-L-and-PsubLIN}(4).

\begin{proofof}{(4)}
Our goal of this proof is to construct an oracle $D$ such that (i) $\psublin^D\nsubseteq \para\dl^D$, (ii) $\para\p^D\subseteq \psublin^D$, (iii) $\para\nl^D\subseteq \para\mathrm{LOG}^2\mathrm{SPACE}^D$, and (iv) $\co\para\nl^D\subseteq \para\nl^D$. A basic idea of constructing $C$ is similar to the one used in Theorem \ref{oracle-L-and-PsubLIN}(2). Cases $l\in\{1,3,4\}$, which target Conditions (i) and (iii)--(iv), are similar to (2). From Condition (i), it follows that $\para\p^D\nsubseteq \para\dl^D$. At Stage $t=(l,i)$, assume that $D' = \bigcup_{t'<t} D_{t'}$ and $R'=\bigcup_{t'<t} R_{t'}$ have been already defined. Let $r_x=\ceilings{\sqrt{|x|}}$.

\s

(a) Case $l=2$. This case is meant to satisfy Condition (ii). We consider $S_n$ together with a linear ordering $<$ with respect to $\{b_i\}_{i\in\nat^{+}}$, as defined before. Take all pairs $(x,i)$ one by one. Let $k_{x,i} = (|x|m_i(x))^{b_i}+b_i$ and consider the machine $P_i$, which runs on any input $x$ in time at most $k_{x,i}$. To simulate $P_i$, we use the following DTM $H_i$. On input $w$, $H_i$ makes queries of the form $1^401^iw\# 0^{k_{w,i}} y_j$ with $y_j = 10^{r_w-j}1^j$ for all $j\in[r_w]$ and collect their oracle answers $u_j = D'(1^401^iw\# 0^{k_{w,i}}y_j)$. Letting $u=u_1u_2\cdots u_{r_w}$, $H_i$ then queries the word $1^401^iw\# 1^{k_{w,i}}u$ to an oracle.  If the oracle answers  YES, then we accept $w$; otherwise, we reject $w$. This machine $H_i$ is indeed an oracle $\psublin$-machine.

If $x$ is the first string in $\Sigma^n$, then we set $D_{t,<(x,i)} = R_{t,<(x,i)} =\setempty$. If $x$ is not the first element, then we set $D_{t,<(x,i)}$ as $\bigcup_{y<(x,i)} D_{t,y}$ and set $R_{t,<(x,i)}$ as $\bigcup_{y<(x,i)} R_{t,y}$. Let $\tilde{R}_{t,x,i} = \{1^401^ix\#1^{k_{x,i}}y_j\mid j\in[r_x]\}$.
Since there is enough room for $D_{t,(x,i)}$ by Case $l=1$, we can choose a set $D_{t,(x,i)} \subseteq \tilde{R}_{t,x,i} \cup \{1^401^ix\#1^{k_{x,i}}u\} - R'\cup R_{t,<(x,i)}$ so that $x\in L(P_i,D'\cup D_{t,<(x,i)}\cup D_{t,(x,i)})$ iff $x\in L(H_i,D'\cup D_{t,<(x,i)}\cup D_{t,(x,i)})$.
Finally, we define $R_t$ to include $\tilde{R}_{t,x,i}\cup D_{t,(x,i)}$ and all query words of $P_i$ as well as $H_i$ on $x$ relative to $D'\cup D_{t,<(x,i)}$ for all pairs $(x,i)\in S_n$.

\s

It is not difficult to show that Condition (iv) is satisfied.
\end{proofof}

Combining all the proofs for (1)--(4), we now complete the proof of Theorem \ref{oracle-L-and-PsubLIN}.

\section{Proof of Theorem \ref{oracle-NL-PsubLIN}}\label{sec:proof-of-Theorem-2}

We will verify Theorem \ref{oracle-NL-PsubLIN}. In Theorem \ref{oracle-L-and-PsubLIN}, we utilize the fact that the inclusion relationship of $\para\dl^A\subseteq \psublin^A\subseteq \para\p^A$ holds for any oracle $A$. Unlike this case, we cannot expect a similar inclusion relationships for $\para\dl^A$, $\para\nl^A$, and $\psublin^A$.
Since the proof of Theorem \ref{oracle-NL-PsubLIN} requires effective enumerations of various oracle machines, we need to recall from Section \ref{sec:proof-of-Theorem-1} the effective enumerations $\{(M_i,m_i)\}_{i\in\nat^{+}}$,  $\{(D_i,m_i)\}_{i\in\nat^{+}}$, $\{(N_i,m_i)\}_{i\in\nat^{+}}$, and $(K_1,K_2,\ldots)$. Moreover, we recall two index sets $\Lambda$ with $\Theta$ and $S_n$ with its linear ordering $<$ from Section \ref{sec:proof-of-Theorem-1}.

\begin{proofof}{(1)--(2)}
(1) This follows directly from Theorem \ref{oracle-L-and-PsubLIN}(1).

(2) We require the following four conditions: (i) $\psublin^B\nsubseteq \para\dl^B$ and (ii) $\para\nl^B\subseteq \para\dl^B$. Note that Condition (ii) implies that $\para\nl^B = \co\para\nl^C = \para\dl^B \subseteq \para\mathrm{LOG}^2\mathrm{SPACE}^B$.
Condition (i) can be dealt with in a way similar to the proof of Theorem \ref{oracle-L-and-PsubLIN}(2).
Condition (ii) is also similar to the proof of Theorem \ref{oracle-L-and-PsubLIN}(3).
\end{proofof}

Next, we prove the third claim of Theorem \ref{oracle-NL-PsubLIN}.

\begin{proofof}{(3)}
Since $\para\dl^{C}\subseteq \psublin^C$ holds for any $C$, $\para\nl^C\nsubseteq \psublin^C$ implies that $\para\dl^C\neq\para\nl^C$. Hence, we demand that the desired oracle $C$ should satisfy that (i) $\para\nl^C\nsubseteq \psublin^C$, (ii)  $\para\nl^C \subseteq \para\mathrm{LOG}^2\mathrm{SPACE}^C$, and (iii) $\co\para\nl^C\subseteq \para\nl^C$.
For this proof, we use an example language $L^C=\{101^ix\mid \exists z\in\Sigma^{|x|} [101^ix\# z\in C]\}$ for an oracle $C$. Note that $(L^C,||)\in \para\nl^C$ for any $C$.

In what follows, we want to construct the desired oracle $C$ by stages. For the construction of $C$, we use $\Theta=\{1,2,3\}$ and define an index set $\Lambda$ as done before.  At each stage, we want to define $C_t$ and also define a set $R_t$ of \emph{reserved words}. We will define a series $\{n_s\}_{s\in\nat^{+}}$ of numbers by stages.

At Stage $0$, we set $n_0=0$, $C_0= R_0=\setempty$ and $n_0=1$. We also prepare a counter $s$, starting at $s=1$. Let us consider Stage $t=(n,l)$ in $\Lambda$ with a counter $s$. We assume that, for all elements $t$ in $\Lambda$ satisfying $t<(n,l)$, the sets $C_t$ and $R_t$ have been already defined. Moreover, we have already defined all $n_e$ for $e<s$.
For brevity, let $C'=\bigcup_{t<(n,l)} C_t$ and $R'=\bigcup_{t<(n,l)}R_t$.
We will describe how to define $C_t$ and $R_t$ depending on the values of $l$ and $n_s$.
Cases $l\in\{2,3\}$, which target Conditions (ii)--(iii), are similar to Theorem \ref{oracle-L-and-PsubLIN}(2). Here, we explain only Case $l=1$. Assume that $s$ and $n_{s}$ have been already defined.

\s

(a) Case $l=1$. In this case, we will target Condition (i). Take a simple size parameter $m$ defined by $m(x)=|x|$ for all $x$. Whenever $n<n_s$, we skip this case. Next, we assume that $n=n_s$. Let $V_{x,n} = \{101^ix\# z\mid z\in\Sigma^{n}\}$. Letting $k_{x,i} = (|x|m_i(x))^{e_i}+e_i$, we define $Z^{(2)}_{x,n} = \{1^201^ix\# 0^{k_{x,i}}z\mid |z|=\ceilings{2\log\log{n}}\}$ and $Z^{(3)}_{x,n} = \{1^301^ix\# z\mid |z|= k_{x,i}\}$.
Check whether there exist a number $\tilde{n}$ and a string  $x\in\Sigma^{\tilde{n}}$ such that (*) $\max_{i\in[\log^*{n}]} \{4n^{2c_ig_i}+c_i\} < \tilde{n}\leq 2^{n_{s}}$ and, for any $j\in\{2,3\}$,  $|R'\cap Z^{(j)}_{x,i}| + |\tilde{x}|^{2e_i}+e_i < \tilde{n}^{\log{\tilde{n}}}$, where $\tilde{x}=101^ix$.
If (*) is not satisfied for all $\tilde{n}$ and $x$, then we skip this case. Hereafter, we assume that (*) holds for certain $\tilde{n}$ and $x\in\Sigma^{\tilde{n}}$. take such a pair $(x,\tilde{n})$. We consider $M_i$, which on input $\tilde{x}$ runs in time at most $|\tilde{x}|^{2e_i}+e_i$ using space at most $|\tilde{x}|^{\varepsilon_i}\log^{k_i}|x|+e_i$.  Note that $M_i$ cannot query all strings in $V_{x,\tilde{n}}-R'$. Our goal is to construct $C_{t}$ (as well as $R_{t}$) such that $\tilde{x}\in L(M_i,C'\cup C_t)$ iff $\tilde{x}\notin L^{C'\cup C_t}$, where $\tilde{x}=101^ix$.
For this purpose, we define $\tilde{R}_t$ to be the set of all queried words of $M_i$ on $\tilde{x}$ relative to $C'$. We also  define $C_t=\{101^ix\# z_{x,\tilde{n}}\}$ with $z_{x,\tilde{n}} =\min \{z\in\Sigma^{\tilde{n}} \mid 101^sx\#z\notin R'\cup \tilde{R}_t\}$ if $M_i$ rejects $\tilde{x}$, and $C_t=\setempty$ otherwise. Define $R_t = \tilde{R}_t \cup C_t$. Before leaving this case, we define $n_{s+1}$ to be $\tilde{n}$ and then increment the counter from $s$ to $s+1$.

\s

By the construction of $C$, Conditions (i)--(iii) are all satisfied.
\end{proofof}


To close the proof of Theorem \ref{oracle-NL-PsubLIN}, we will verify the fourth claim of the theorem.

\begin{proofof}{(4)}
Hereafter, we want to show that, for a certain recursive oracle $D$, (i) $\para\nl^D\nsubseteq \para\dl^D$, (ii)  $\para\nl^D\subseteq \para\mathrm{LOG}^2\mathrm{SPACE}^D$, (iii) $\co\para\nl^D \subseteq \para\nl^D$, and (iv) $\para\nl^D \subseteq \psublin^D$.

Stage by stage, we will construct the desired oracle $D$. We use the index set $\Lambda$.  Cases $l\in\{2,3\}$, which respectively correspond to Conditions (ii)--(iii), are similar to the ones in the proof of Theorem \ref{oracle-L-and-PsubLIN}(2). Hence, we will target Conditions (i) and (iv).
We will define $\{n_s\}_{s\in\nat^{+}}$, $\{D_t\}_{t\in\Lambda}$, and $\{R_t\}_{t\in\Lambda}$ by stages. At Stage $t=(l,i)$, we assume that all stages $t'<(n,l)$ have already been processed. Let $D'= \bigcup_{t'<t} D_{t'}$ and $R' = \bigcup_{t'<t}R_{t'}$. Let $k_{x,i} = (|x|m_i(x))^{e_i}+e_i$. We use $L^D = \{101^ix\mid \exists z\in\Sigma^{|x|} [ 101^ix\# z\in D]\}$ as an example language. We define $Z^{(2)}_{x,n} = \{1^201^ix\# 0^{k_{x,i}}z \mid |z|=\ceilings{2\log\log{n}}\}$, $Z^{(3)}_{x,n} = \{1^301^ix\# z \mid |z|= k_{x,i}\}$,
 $Z^{(4)}_{x,n} = \{1^401^ix\# ^{k_{x,i}}y_j \mid y_j=10^{r_x-j}1^{j},j\in[r_x]\}$, where $r_x=\ceilings{\sqrt{|x|}}$.

\s

(a) Case $l=1$. This case corresponds to Condition (i) and it can be handled in a way similar to (a) of the proof of (3). Let $s$ be the value of a counter. If $n<n_s$ holds, then we skip this case and move to case $l=2$. Let us assume that $n=n_s$. Check if there is a pair of $\tilde{n}$ and $x\in\Sigma^{\tilde{n}}$ satisfying that (*) $\max_{i\in[\log^*{n}]}\{4n^{c_ig_i}+c_i\}< \tilde{n}\leq 2^{n_s}$ and, for any index $j\in\{2,3,4\}$,  $|R'\cap Z^{(j)}_{x,\tilde{n}}|+ |\tilde{x}|^{2a_i}+a_i <2^{\sqrt{\tilde{n}}}$. Let $V_{x,\tilde{n}} = \{101^ix\# z\mid z\in \Sigma^{\tilde{n}}\}$.
If (*) is not satisfied for all pairs $(\tilde{n},x)$, then we skip this case and advance to Case $l=2$. Next, we assume that (*) holds for a certain pair $(\tilde{n},x)$. Fix such a pair $(x,\tilde{n})$. Let us consider $L^D$ and $M_i$ running on the input $\tilde{x}$. Let $\tilde{R}_t$ be composed of all query words of $D_i$ on $\tilde{x}$ relative to $D'$. It follows that $|\tilde{R}_t|\leq |\tilde{x}|^{2a_i}+a_i$ because of the runtime of $M_i$. We choose the lexicographically smallest string $z_{x,i}$ in $\Sigma^{|\tilde{x}|}$ satisfying $101^ix\# z\notin R'\cup \tilde{R}_t$ and define $D_t = \{101^ix\# z_{x,i}\}$ if $D_i$ rejects $\tilde{x}$, and $D_t=\setempty$ otherwise. Finally, we set $R_t = \tilde{R}_t\cup V_{x,\tilde{n}}$. We also set $n_{s+1}=\tilde{n}$ and update the counter from $s$ to $s+1$.

\s

(b) Case $l=4$. We aim at Condition (iv). Assume that $D'$ and $R'$ have been already updated. Recall an oracle $\psublin$-machine $H_i$ from the proof of Theorem \ref{oracle-L-and-PsubLIN}(4). Here, we use $S_n$ and its linear ordering $<$ with respect to $\{e_i\}_{i\in\nat^{+}}$. We pick up all pairs $(x,i)$ in $S_n$ one by one in order and assume that $D_{t,<(x,i)}$ and $R_{t,<(x,i)}$ have been defined.
The machine $H_i$ in the proof of Theorem \ref{oracle-L-and-PsubLIN}(4) makes queries of the form $1^401^ix\# 0^{k_{x,i}} y_j$ with $y_j = 10^{r_x-j}1^j$ for all $j\in[r_x]$, where  $r_x=\ceilings{\sqrt{|x|}}$. Collect their oracle answers $u_j = D'(1^401^ix\# 0y_j)$. Finally, query the word $1^401^ix\# 1^{k_{x,i}} u$, where $u=u_1u_2\cdots u_{r_x}$. If its oracle answer is YES, then accept $x$; otherwise, reject $x$. We define $D_{t,(x,i)}$ as follows.
If $N_i$ rejects $x$ relative to $D'\cup D_{t,<(x,i)}$, then we choose a set  $D_{t,(x,i)} \subseteq \{1^401^ix\# 0^{k_{x,i}} y_j\mid j\in[r_x]\}\cup \{1^401^ix\# 1^{k_{x,i}} u\} - R'\cup R_{t,<(x,i)}$, where  $u=u_1u_2\cdots u_{r_x}$ and $u_j= D'(1^401^ix\# 0^{k_{x,i}} y_j)$ for all $j\in[r_x]$, such that $x\in L(N_i,D'\cup D_{t,<(x,i)}\cup D_{t,(x,i)})$ iff $x\in L(H_i,D'\cup D_{t,<(x,i)}\cup D_{t,(x,i)})$. Otherwise, we set $D_{t,(x,i)}=\setempty$. Since  $|R'\cap \{1^401^ix\# z\mid z\in\Sigma^*\}|\leq |\tilde{x}|^{2a_i}+a_i < 2^{\sqrt{n}}$ by (a), $D_{t,(x,i)}$ must exist. Before leaving this case, we set $R_{t,(x,i)}$ to be the set of all query words of $N_i$ and $H_i$ on $x$.

\s

Therefore, the construction of $D$ for Conditions (i)--(iv) are satisfied.
\end{proofof}

\section{A Brief Discussion on Supportive Oracles}\label{sec:discussion}

We have introduced the notion of ``$\nl$-supportive oracle''
to guarantee the known inclusion relationships associated with $\para\nl$ and thus make the relativization of $\para\nl$ keep its validity and meaningfulness  in providing good information on the structural similarities and differences between $\para\nl$ and other parameterized complexity classes. With this notion, we have been able to demonstrate the existence of various relativized worlds in which different inclusion and separation relationships occur among four parameterized complexity classes: $\para\dl$, $\para\nl$, $\psublin$, and $\para\p$.

The usefulness of ``supportive oracles'' can be justified by the following argument. Take a quick look at a longstanding open problem: the $\p=?\np$ problem.  There are known recursive oracles $A$ and $B$ for which $\p^A=\np^A$ and $\p^B\neq\np^B$ \cite{BGS75}. Once either $\p=\np$ or $\p\neq\np$ is proven in the unrelativized world, the currently known relativization methodology produces a contradicting result against either $\p=\np$ or $\p\neq\np$. Therefore, we no longer use the current definition of relativization of $\p$ and $\np$ for a further study on relativized worlds associated with $\p$ and $\np$. However, if we consider ``$\np$-supportive oracles,'' which supports the correct relativized relationship of either $\p^A=\np^A$ or $\p^A\neq \np^A$, depending on either $\p=\np$ or $\p\neq\np$, then we cannot construct any conflicting relativized world and it thus remains worth investigating the relativization of complexity classes in relation to $\p$ and $\np$.

We strongly hope that the notion of supportive oracle will prove its importance in computational complexity.

\let\oldbibliography\thebibliography
\renewcommand{\thebibliography}[1]{%
  \oldbibliography{#1}%
  \setlength{\itemsep}{0pt}%
}
\bibliographystyle{plainurl}

\end{document}